\documentclass[10pt,showpacs,preprintnumbers,footinbib,amsmath,amssymb,aps,prl,twocolumn,groupedaddress,superscriptaddress,showkeys]{revtex4-1}
\usepackage{graphicx}
\usepackage{dcolumn}
\usepackage{bm}
\usepackage[colorlinks=true,urlcolor=blue,citecolor=blue]{hyperref}
\usepackage{color}

\begin{document}

\preprint{LLNL-JRNL-701385}

\title{Realistic calculations for $c$-coefficients of the isobaric
mass multiplet equation in $1p0f$ shell nuclei}

\author{W.~E.~Ormand}
\affiliation{Lawrence Livermore National Laboratory, Livermore, CA 94551, USA}
\author{B.~A.~Brown} 
\affiliation{Department of Physics and Astronomy and National Superconducting Cyclotron Laboratory,
Michigan State University, East Lansing, MI 48824-1321, USA}
\author{M.~Hjorth-Jensen}
\affiliation{Department of Physics and Astronomy and National Superconducting Cyclotron Laboratory,
Michigan State University, East Lansing, MI 48824-1321, USA}
\affiliation{Department of Physics, University of Oslo, N-0316, Oslo, Norway}

\begin{abstract}
We present calculations for the $c$-coefficients of the isobaric
mass multiplet equation
for nuclei from $A=42$ to $A=54$ based on input from three 
realistic nucleon-nucleon interactions.
We demonstrate that there is a clear dependence on the short-ranged charge-symmetry
breaking (CSB) part of the strong interaction and that there is significant
disagreement in the CSB part between the commonly
used CD-Bonn, N$^3$LO, and Argonne V18 nucleon-nucleon
interactions. In addition, we show that all three interactions give a CSB contribution 
to the $c$-coefficient that is too large when compared to experiment.
\end{abstract}

\pacs{21.30.Fe, 21.60.Cs, 21.60.De, 27.40.+z}

\maketitle

Isospin is a powerful spectroscopic tool in nuclear physics that can
be used to label and characterize states not only in a specific
nucleus, but also corresponding states in an analog nucleus. Isospin,
denoted by $T$, is an additive quantity similar to the intrinsic spin
of the proton and neutron~\cite{heisenberg}. The charge, $Q$, of 
the particle is defined by the $z$-component via $Q=\frac{1}{2} + T_z$. 
Thus, a nucleus with $Z$ protons and $N$ neutrons has 
 $T_z=(Z-N)/2$ and may have 
isospin states with $T_z \le T \le (Z+N)/2$. Isospin symmetry is broken 
by components in the nuclear Hamiltonian that treat
protons and neutrons differently. The most obvious, and significant,
component is the Coulomb interaction acting only between protons due
to their electric charge. There are, however, weaker isospin-symmetry
breaking components in the nucleon-nucleon interaction itself caused
by differences in the masses of up and down quarks and their intrinsic
electric charges, which is reflected in the slightly different masses
exhibited by neutrons and protons \cite{miller2006} and the slightly
different strong-interaction scattering lengths observed in the
proton-proton ($pp$), neutron-neutron ($nn$), and the $T=1$
proton-neutron ($pn$) channels~\cite{henley1969,bergervoot1988,
cdbonn2001,trotter2006,chen2008}.

Important signatures of isospin-symmetry breaking interactions are
differences in the binding energy of nuclei within the same isospin
multiplet with fixed nucleon number $A$. These mass splittings, or
Coulomb-displacement energies, offer a sensitive probe of the
properties of isospin-symmetry breaking in nuclei. The three $T=1$
nucleon-nucleon channels can be decomposed into three isospin
components: isoscalar (rank 0), isovector (rank 1), and isotensor
(rank 2), defined in terms of the $pp$, $nn$, and $pn$ interactions
via
\begin{align}
\label{e:1}
v^{(0)} & =  \frac{1}{3} (v_{pp} + v_{nn} + v_{pn}) \\
\label{e:2}
v^{(1)} & =  (v_{pp} - v_{nn}) \\
\label{e:3}
v^{(2)} & =  v_{pn} - \frac{1}{2}(v_{pp} + v_{nn}).
\end{align}
With these three components, the masses for a set of states within a multiplet 
with isospin $T$ may be described by the isobaric mass multiplet 
equation (IMME)~\cite{wigner1957}
\begin{equation}
M (T_z) = a + bT_z + cT_z^2,
\end{equation}
where the coefficients $a$, $b$, and $c$ are dependent on the
isoscalar, isovector, and isotensor components of the nuclear
Hamiltonian, respectively. The linear and quadratic dependence on
$T_z$ is due to the application of the Wigner-Ekhart theorem and the
appropriate Clebsch-Gordon coefficients arising for the isovector and
isotensor components of the Hamiltonian, respectively. For $T=1$ states, 
the $b$- and $c$-coefficients are equivalent to half the 
mirror-energy displacement (MED) and triple-energy displacement (TED), respectively,
discussed in Refs.~\cite{zuker2002, gadea2006}. In these two references, the 
angular momentum, or $J$-dependence of these quantities; where it was 
concluded that the observed $J$-dependence in the MED and TED was
explained within the context of the underlying two-body matrix elements (TBME),
and that overall, an empirically determined correction relative to the Coulomb 
TBME was needed. In earlier empirical studies of isospin-nonconserving 
interactions for $0s1d$ and $1p0f$-shell nuclei, it was found that globally $c$-coefficients
are well reproduced if the $T=1$ $pn$ interaction is 2\% more attractive than the average
of $pp$ and $nn$~\cite{ormand_inc}.

Here, we compute $c$-coefficients (TED) as a function
of excitation energy and angular momentum for nuclei in the mass range 
$42 \le A \le 54$ using the Coulomb interaction and isospin-symmetry 
breaking interactions derived from three realistic nucleon-nucleon 
interactions utilizing well-known renormalization 
procedures~\cite{mhj1995}. We calculate the effect of charge-symmetry breaking in the strong force
on the $c$-coefficients and demonstrate that effective two-body CSB interactions derived 
from state-of-the art nucleon-nucleon interactions each fail to describe experimental 
data. Further, we demonstrate that at this level, that the CSB interactions derived from the three 
realistic interactions are in significant disagreement with each other.
This signifies either: 1) a deficiency in our understanding 
of isospin-symmetry breaking in the nucleon-nucleon interaction, 
2) significant isospin-symmetry breaking in the initial three-nucleon interaction, 
or 3) large contributions to isospin-symmetry breaking in three-nucleon 
interactions induced by the renormalization procedure. 

We performed a series of shell-model calculations using the program
BIGSTICK~\cite{bigstick,shan2015} to compute the $c$-coefficients of the IMME for odd-odd 
$N=Z$ nuclei and their $T=1$ analogs in the $1p0f$ shell with $ 42 \le $ A $ \le 54$. 
Calculations were performed with the full $1p0f$-shell model 
space, except for $A=54$ where up to five particles excited from the $0f_{7/2}$ orbit were 
permitted with $M$-scheme dimensions $\sim$ 500 M~\cite{truncation}. 
The $c$-coefficients were computed with CSB-interactions derived from each realistic 
nucleon-nucleon interactions using renormalization techniques and many-body 
perturbation theory as described in Ref.~\cite{mhj1995}. 
The two-body matrix elements were computed in
two steps. In the first step, the nuclear two-body interaction was renormalized 
using either the $G$-matrix approach~\cite{bethe1963,day1967,bethe1971}  
or the V$_{low~k}$ method~\cite{nogga2003}; both schemes give almost 
indistinguishable  effective interactions. The second step consisted in
obtaining an effective interaction tailored to a small shell-model
space using many-body perturbation theory up to
$3^{\rm rd}$ order with the renormalized nucleon-nucleon interaction, which 
includes the so-called folded diagrams \cite{mhj1995}. All codes used to
generate these interactions are publicly available~\cite{mhjgit}.  

To derive the nuclear CSB interactions, we employed the realistic
N$^3$LO \cite{entem2003}, AV18 \cite{argonne1995} and CD-Bonn
\cite{cdbonn2001} nucleon-nucleon interactions. These interaction
models include breaking of isospin symmetry and charge symmetry in the strong
interaction.  We note that the AV18 interaction also includes detailed 
electromagnetic corrections and the full interaction potential was used 
in the first step, whereas Coulomb was included for the N$^3$LO and 
CD-Bonn interactions after renormalization.
The two-body matrix elements of the Coulomb and nucleon-nucleon 
interactions were computed using a harmonic oscillator (HO) 
basis with an oscillator energy
$\hbar\omega =10.5$ MeV with an effective Hilbert space defined by the
first twelve oscillator shells. The V$_{low~k}$ interactions were
obtained with a cut-off parameter of $\Lambda$ = 2.1 fm$^{-1}$.  The
model-space effective interaction was computed with and without the Coulomb
interaction, and the Coulomb two-body matrix elements were obtained
from the difference between these proton-proton ($pp$) matrix
elements. The renormalized interaction computed without Coulomb was
then decomposed into the three isospin components: isoscalar (rank 0),
isovector (rank 1), and isotensor (rank 2), as defined in
Eqs.~(\ref{e:1})-(\ref{e:3}). 

The validity of the use of harmonic-oscillator radial wave functions 
for the Coulomb interaction was tested by 
performing an energy-density functional (EDF) calculation for $^{48}$Cr 
with the SkX Skyrme functional~\cite{ref:SkX}. From this, we obtained 
the $\hbar\omega$ needed to reproduce the calculated {\it rms} charge radius 
(10.72 MeV). We then calculated the Coulomb two-body matrix
elements (TBME) with the EDF and HO radial wave functions for the
$1p0f$ orbitals. The average difference for the diagonal TBME for all
orbitals was about 1 keV. The difference for the most important 
$0f_{7/2}$ orbital was (13,10,8,8) keV for $J$ = (0,2,4,6) TBME. 
For our application, we conclude that it is sufficient to use the HO basis for 
the Coulomb matrix elements as long as $\hbar\omega$ is scaled 
according to the total {\it rms} radius. Here, the $A$-dependence was 
properly accounted for by scaling the Coulomb matrix elements by 
$ \sqrt{\hbar\omega(A)/10.5} $, where $\hbar\omega(A)$ was 
determined from the {\it rms} radius obtained from a spherical EDF 
calculation for $42 \le A\le 54$ nuclei using the SkX Skyrme functional. 

The $c$-coefficients of the IMME were obtained utilizing first-order
perturbation theory. The base for each calculation was the eigenstate,
$E_{0}$, for each member of the $T=1$ triplet, $ \vert T_{z} \rangle$,
obtained using the isoscalar GX1A Hamiltonian~\cite{ref:GX1A}. The
GX1A interaction was used instead of the $v^{(0)}$ interaction
obtained from the realistic interaction described above because of
well-known extensions that must be included to properly capture the
behavior of higher-order components and the three-body interaction in
the traditional configuration-interaction shell model for atomic
nuclei, see for example Refs.~\cite{zuker2003,ekstrom2015}. The 
TBME for the derived isovector and isotensor interactions were 
assumed to have the same $A$-dependence as the GX1A 
interaction. The expectation value of the Coulomb,
isovector, and isotensor interactions are then computed to give the
full energy for each state,
\[
E(T_{z}) = E_{0} + \langle T_{z} \vert   v^{{\rm Coul}} + v^{(1)} + v^{(2)}\vert T_{z} \rangle
\]
the c-coefficient is then computed from
\[
c = [E(T_{z} =1) - 2E(T_{z} =0) + E(T_{z} =-1)]/2.
\]

Figure 1 shows the typical dependence on the order of many-body perturbation theory as demonstrated by the CD-Bonn interaction. In the right-hand panel, the contribution from 
Coulomb is shown for each order. In the left-hand panel, the dashed lines show the CSB contribution from the CD-Bonn interaction, while the solid lines show the full value obtained by adding the Coulomb and CSB components for each order. The figure demonstrates that the 
$J$-dependence of the Coulomb- and CSB-contributions is quite
different. The long-range Coulomb has a relatively flat $J$-dependence
with only a small rise at $J=0$. On the other hand, the CSB contribution at $A=42$ shows
a peak at $J=0$ with a sharp drop towards $J=2$, which is
characteristic of a short-ranged interaction.  This same pattern is also observed with a simple $\delta$-function interaction model and the empirical CSB interaction in Refs.~\cite{zuker2002,ormand_inc}

For $A=42$, $J=6$ is
the maximum angular momentum (for $T=1$) in the $1p0f$ model space.
For higher values of $A$, this sharp drop at $ J=2 $ is replaced by a
linear drop to $J=6$ due to configuration mixing.  We note that for
$J=8$ and $10$, the effect of charge-symmetry breaking is small. The
experimental data is taken from the compilation \cite{2013la}, except
for $A=46$, where we use the results from Fig.~2 of \cite{2001ga}.

Both Coulomb and CSB have a small increase at $J=12$.
The reason for this is that
protons with $J=6$ and neutrons with $J=6$
are maximally aligned, resulting in an enhancement of the
overlapping proton and neutron density distributions.

\begin{figure}
\includegraphics[scale=0.4]{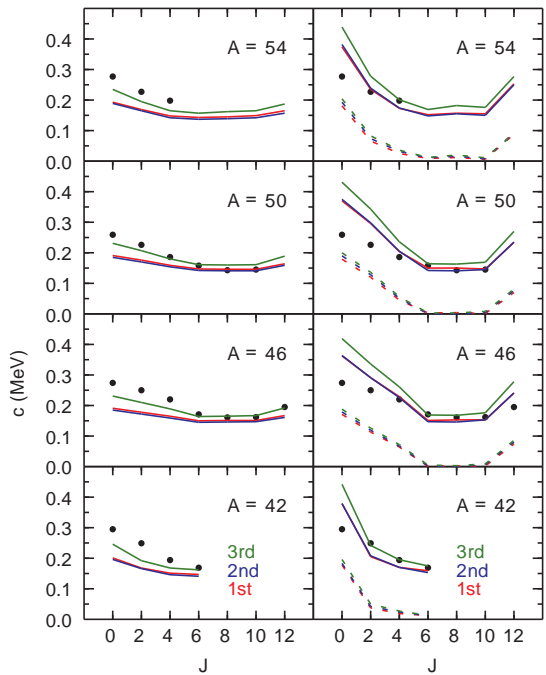}
\caption{(color online) Results for the CD-Bonn potential
up to 1$^{\rm st}$ (red), 2$^{\rm nd}$ (blue), and 3$^{\rm rd}$ (green) order.
The black circles
are the experimental data. The solid lines in the right-hand panel show the Coulomb
contribution to the $c$-coefficients. In the left-hand panel, the dashed lines show the CSB contribution from CD-Bonn, 
while the full line represents the full calculation, CSB + Coulomb.}
\end{figure}
\begin{figure}
\includegraphics[scale=0.4]{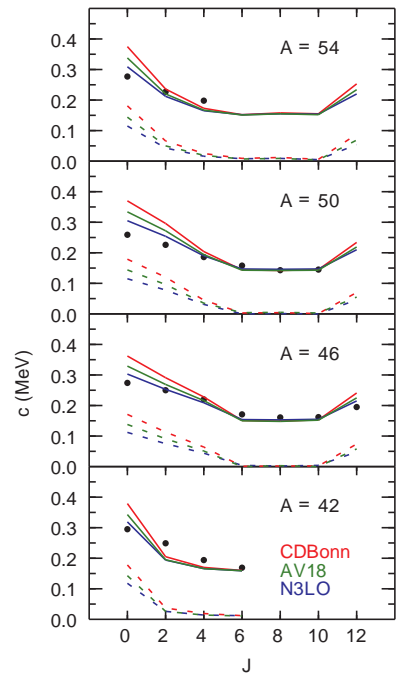}
\caption{(color online) first-order calculations compared to
experiment. The black circles
are the experimental data. The solid lines show the sum of Coulomb and
CSB contributions. The dashed lines show only the CSB contribution.}
\end{figure}
\begin{figure}
\includegraphics[scale=0.4]{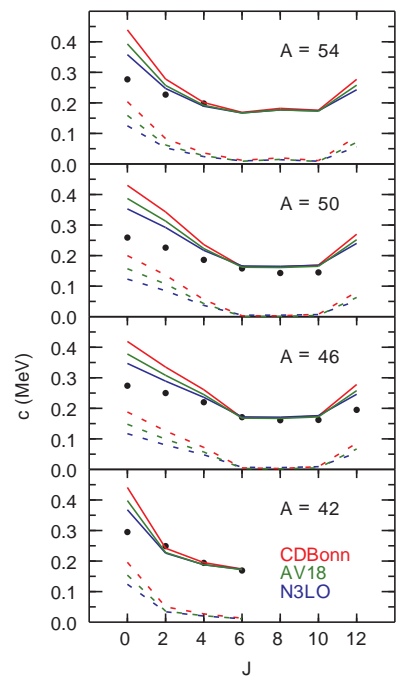}
\caption{(color online) Calculations up to 3$^{\rm rd}$ order compared to
experiment. See caption to Fig 2.}
\end{figure}
\begin{figure}
\includegraphics[scale=0.4]{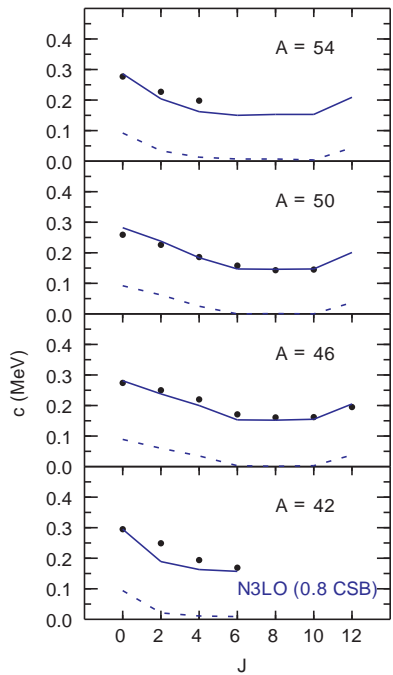}
\caption{(color online) first-order calculations for N$^3$LO
with the CSB part multiplied by 0.8 and compared to
experiment. See caption to Fig 2.}
\end{figure}

The CSB contribution turns out to be almost order independent, while
the Coulomb contribution is almost the same at $1^{\rm st}$ and
$2^{\rm nd}$ order in many-body perturbation theory, but increases by 10-20\% at $3^{\rm rd}$ 
order. This suggests that the CSB interaction is substantially
short-ranged in nature, and the $G$-matrix and $V_{low~k}$ treatment
may be sufficient. A simple analysis of all $J=0$ two-body matrix elements, using
Eqs.~(\ref{e:1})-(\ref{e:3}), shows that for the core-polarization 
contribution at $2^{\rm nd}$ order, the 
correction to $c$-coefficient is about ten times smaller than that for the $a$-coefficient. 
This applies to most two-body matrix elements 
that define Eqs.~(\ref{e:1})-(\ref{e:3}).

It is remarkable that the experimental data are in
rather good agreement with the third-order Coulomb result,
where there seems to be no need for CSB even though this component is
well known to be important in nucleon-nucleon ($NN$) scattering data that is incorporated
into the potential models.

Figure 2 shows the results for the three potential models 
to $1^{\rm st}$ order in many-body perturbation theory. 
This shows that the CSB contribution is model
dependent.  There could be a few reasons for this. While the $NN$
interactions are all fit to scattering data and reproduce the
nucleon-nucleon scattering length equally well, there could be
differences in the underlying treatment of the CSB components.  
For example, while AV18 is a purely local potential, both N$^3$LO and
CD-Bonn are non-local, albeit in different ways. The
short-range correlation effects taken into account in the $G$-matrix and V$_{low~k}$
renormalizations could have different effects on this small component
of the $NN$ interactions, which may be corrected when induced
three-nucleon terms are included. We note that the results obtained
with N$^3$LO are in best agreement with experiment, although all three
interactions over predict the $c$-coefficients. The fact that all three interactions
significantly over predict experiment might also be an indication of 
charge-symmetry breaking in the initial three-nucleon interaction.

Figure 3 shows the results for the three potentials at $3^{\rm rd}$ order.

Our work suggests future investigations to discover the full extent of the 
nature of charge symmetry breaking in the nuclear force.  For better first
principles calculations, one should understand the origin of the
different CSB contributions from these three realistic potentials.  In
particular, in the spirit of using nuclear data to constrain the $NN$
and three-nucleon ($3N$) interactions (in addition to $NN$ scattering data) one should use
the $c$-coefficient as a constraint on the CSB part. From a practical
point of view, we start with the fact that first-order Coulomb
plus CSB is already close to the data.  We can make it almost perfect
by taking the first-order Coulomb contribution and add 80\% of the N$^3$LO
CSB part. This is shown in Fig.~4.  

The largest deviation between our calculations and 
experiment is for $ J=2 $ in $A=42$.  
In $A=42$, the experimental data for $J=0-6 $ fall off in a manner that is similar to that exhibited in the calculation for $A=46$,
while the calculated fall off is more similar to that calculated for $A=54$. 
This is explained by noting that 
the theoretical wave functions for $A=42$ are dominated by $(0f_7/2 )^2$
configurations, while the wave functions for $A = 54$ are
dominated by $(0f_7/2 )^{-2}$ configurations, thus, the calculated $J$-dependence 
between $A=42$ and 54 is similar. However, the experimental $J$ dependence for
$A=42$ is far more similar to that of $A=46$. The reason for this is
that $A=42$ is not well described in the $1p0f$ model space alone due to mixing
with configurations involving nucleons excited from $1s0d$ orbits to the 
$1p0f$ shell, as is exhibited by the fact that the B(E2) value for the 
$J=2^+ \rightarrow 0^+$ transition in $^{42}$Ca is about 10 times larger than 
that calculated in the $1p0f$ model space~\cite{brown1977}. 

The overprediction of the $c$-coefficient for CD-Bonn was also noted in  
Ref.~\cite{caurier2002} which performed {\it ab initio} calculations for $A=10$ nuclei
within the framework of the No-core Shell Model using the
CD-Bonn interaction. The calculated $c$-coefficient was 535 kev, which is
substantially larger than the experimental value of 362 keV.

In conclusion, we have presented the first calculations for the
$c$-coefficients of the IMME for nuclei from $A=42$ to $A=54$, based
on input from three state-of-the-art realistic nucleon-nucleon interactions and their
pertinent shell-model effective interactions. The CSB contribution is
almost independent of the order of renormalization in many-body
perturbation theory, suggesting that the charge-symmetry breaking part
of the interaction is, to a large extent, short-range in nature.  In
effective field theory, this might indicate two-pion, or even higher-order
excitations that probe the short-range nature of the CSB interaction. 
In addition, we find that the three state-of-the art interactions yield 
different results, and are in disagreement. This suggests that either:  1) 
the charge-symmetry breaking in the nucleon-nucleon interaction is
poorly known, 2) there is strong charge-symmetry breaking in the 
three-nucleon interaction, or 3) there are significant induced three-nucleon 
interaction arising from the renormalization procedure.

B.~A.~B. acknowledges U.S. NSF Grant No. PHY-1404442. M.~H.~J. acknowledges
U.S. NSF Grant No. PHY-1404159 and the Research Council of Norway
under contract ISP-Fysikk/216699. W.~E.~O. acknowledges support from the
U.S. Department of Energy, Office of Science, Office of Nuclear
Physics, under Field Work Proposal No. SCW0498. This work was
performed under the auspices of the U.S. Department of Energy by
Lawrence Livermore National Laboratory under Contract
DE-AC52-07NA27344.  Computing support for this work came from the
Lawrence Livermore National Laboratory (LLNL) institutional Computing
Grand Challenge program.


\begin{thebibliography}{99}
\bibitem{truncation} The effect of this truncation on the $c$-coefficients is generally small. Differences between various truncations 
leading to the one implemented were $<$ 5 keV.
\end{thebibliography}
\end{document}